\documentclass[aps,pra,preprint,groupedaddress,superscriptaddress,showpacs]{revtex4-1}
\usepackage{CJK}
\usepackage{amsmath}
\usepackage{longtable}
\usepackage{mathrsfs}
\usepackage{multirow}
\usepackage{tabularx}
\usepackage{array}
\usepackage{graphicx}
\usepackage{bm}

\begin{document}

\title{\textit{Ab initio} calculations of the hyperfine structure of $^{109}$Cd, $^{109}$Cd$^+$ and reevaluation of the nuclear quadrupole moment $Q$($^{109}$Cd)}


\author{Benquan Lu}
\affiliation{National Time Service Center, 710600 Lintong, China}
\affiliation{The University of Chinese Academy of Sciences, 100088 Beijing, China}
\author{Xiaotong Lu}
\affiliation{National Time Service Center, 710600 Lintong, China}
\affiliation{The University of Chinese Academy of Sciences, 100088 Beijing, China}
\author{Tao Wang}
\email{tauwaang@cqu.edu.cn}
\affiliation{Department of Physics, and Center of Quantum Materials and Devices, Chongqing University, Chongqing, 401331, China}
\author{Hong Chang}
\email{changhong@ntsc.ac.cn}
\affiliation{National Time Service Center, 710600 Lintong, China}
\affiliation{The University of Chinese Academy of Sciences, 100088 Beijing, China}


\date{\today}

\begin{abstract}
Large scale \textit{ab initio} calculations of the electric contribution (i.e. the electric field gradient) to the electric quadrupole hyperfine interaction constants $B$ for the $5p~^2P_{3/2}$ state in $^{109}$Cd$^+$ ion and the $5s5p~^3P_{1,2}$ states in neutral $^{109}$Cd atom were performed. To probe the sensitivity of $B$ to different electron correlation effects, six sets of variational multi-configuration Dirac-Hartree-Fock and relativistic configuration interaction calculations employing different strategies were carried out. The calculated electric field gradients, together with experimental values of $B$, allow us to extract a new value of the nuclear electric quadrupole moment $Q$($^{109}$Cd) = 0.634(22)~b which is about 5\% larger than the recommended value [Stone, At. Data Nucl. Data Tables 111-112, 1 (2016); Pyykk\"{o}, Mol. Phys. 116, 1328 (2018)]. Efforts were made to provide a realistic theoretical uncertainty for the final $Q$ value based on statistical principles and correlation with the magnetic dipole hyperfine interaction constant $A$.
\end{abstract}


\maketitle
\section{INTRODUCTION}
Cadmium (Z = 48), which has an open shell and two protons less than ``magic" tin, is of broad interest since its nucleus is closer to the shell and has more characteristics. Furthermore, the knowledge of nuclear properties in the vicinity of the tin isotopes is helpful for understanding stellar nucleosynthesis. In comparison to the magnetic dipole moment $\mu_I$, the nuclear electric quadrupole moment $Q$ can give a more rigorous test of the nuclear shell model since it has a very typical linear behavior with regard to the number of valence nucleons~\cite{Mayer1955,Horie1955}. In addition, the quadrupole moment is a unique and excellent tool to study nuclear deformation and shape coexistence~\cite{0034-4885-66-4-205}, especially for exotic nuclei in the vicinity of the proton or neutron drip line~\cite{Campbell2016127}. The nuclear electric quadrupole moment of cadmium has been proved to be the most revealing case to study the above phenomenons~\cite{Yordanov2013,Yordanov2016}. Additionally, specific questions concerning the nuclear structure of the cadmium isotopes, such as shell quenching~\cite{Kautzsch2000,PhysRevLett.91.162503}, sphericity~\cite{PhysRevLett.100.072501}, deformation~\cite{Jungclaus2006,Rodriguez2008}, or whether vibrational nuclei exist at all~\cite{Garrett2010}, require the nuclear electric quadrupole moment.

The nuclear trends of the nuclear quadrupole moments $Q$ among 12 neutron-rich isotopes in the range $^{107-129}$Cd were studied~\cite{Yordanov2013}. The $Q$(Cd) values given in Ref.~\cite{Yordanov2013} were based on measured atomic hyperfine structures for odd neutron number $N$ isotopes from the collinear laser spectroscopy at ISOLDE-CERN. They were derived by combining the measured hyperfine parameter $B$ of $5p~^2P_{3/2}$ state in cadmium ion with the calculated electronic contribution $B/Q$ for each isotope. The $B/Q$ values were resulted from the Dirac-Hartree-Fock calculation. Afterwards, the nuclear quadrupole moments of $^{107,109,121,123}$Cd and $^{111m-123m}$Cd were obtained from the ratio of the hyperfine interaction constant $B$, $^AQ$/$^{109}Q$ = $^AB$/$^{109}B$, by using the quadrupole moment of the long-lived $^{109}$Cd isotope as a reference~\cite{Froemmgen2015}. The nuclear quadrupole moment of $^{109}$Cd was derived from the combination of the measured $B$ value of $5s5p~^3P_2$ state in neutral $^{109}$Cd atom and the value of $B/Q$. The $B/Q$ was calculated, with an uncertainty of 6\%, using the multi-configuration Dirac-Hartree-Fock (MCDHF) theory, in which the main first-order electron correlations were considered. However, it has been found that it is essential to take into account the higher-order electron correlations for achieving satisfactory accuracy of physical quantities concerned, for example, the hyperfine structure constants for the $5s5p~^1P_1$ and $^3P_{1,2}$ states in $^{87}$Sr~\cite{PhysRevA.100.012504}. In addition, due to the sensitivity of the hyperfine parameters to different electron correlation effects, it is sometimes necessary to do multiple sets of computations, each using a different computing strategy and correlation model, in order to assess the quality of the calculations~\cite{PhysRevA.97.062505}. As a result, to obtain high accurate value of $B/Q$, different computational models should be established to systematically consider the electron correlation effects.

In this work, we performed six different computational approaches to obtain accurate values on the $B/Q$s for the $5p~^2P_{3/2}$ state in $^{109}$Cd$^+$ ion and the $5s5p~^3P_{1,2}$ states in neutral $^{109}$Cd atom by using the MCDHF theory. In our calculation, the electron correlation effects, especially the core-core and higher-order correlations, on the $B/Q$s were investigated in detail. Combining the calculated $B/Q$s with the available experimental $B$, a value of the nuclear quadrupole moment of $^{109}$Cd was evaluated with an uncertainty of 3\%. The uncertainty was estimated based on statistical principles and the error bar of the magnetic dipole hyperfine interaction constant $A$.
\section{Thoery}
In the MCDHF method, the wave function $\Psi (\Gamma PJM_J)$ for an atomic state is expressed as a linear combination of configuration state functions (CSFs) $\Phi_j (\gamma_j PJM_J)$ with the same parity $P$, total angular momentum $J$ and its projection quantum numbers $M_J$, i.e.,
\begin{equation}\label{eq:1}
\Psi (\Gamma PJM_J)=\sum_{j}^{N}c_j\Phi_j (\gamma_jPJM_J),
\end{equation}
where, $c_j$ represents the mixing coefficient corresponding to the $j$th configuration state function, and $\gamma$ stands for the other quantum numbers which can define the state uniquely. The configuration state functions $\Phi_j(\gamma_jPJM_J)$ are built from sums of products of the one-electron Dirac orbitals
\begin{equation}\label{eq:2}
\phi(r,\theta ,\varphi ,\sigma)=\frac{1}{r}\binom{P(r)\chi_{\kappa m}(\theta ,\varphi ,\sigma )}{iQ(r)\chi_{-\kappa m}(\theta ,\varphi ,\sigma )},
\end{equation}
where $P(r)$ and $Q(r)$ are the radial wavefunctions. The coefficients $c_j$ and the radial parts of the one-electron orbitals are optimized simultaneously in the self-consistent field (SCF) procedure. The SCF radial equations to be iteratively solved are derived from the application of the variational principle on a weighted Dirac-Coulomb energy functional of the targeted atomic states according to the extended optimal level (EOL) scheme~\cite{Grant1980}. Once a radial orbital set has been determined, the relativistic configuration interaction (RCI) computation in which only the expansion coefficients are varied. The Breit interaction and quantum electrodynamic (QED) corrections can be included in the RCI computation.

The hyperfine structure of atomic energy levels is caused by the interaction between the electrons and the electromagnetic multipole moments of the nucleus. The interaction Hamiltonian can be represented by a multipole expansion
\begin{equation}\label{eq:3}
H_{hfs}=\sum_{k\geq 1}\mathbf{T}^{(k)}\cdot \mathbf{M}^{(k)},
\end{equation}
where $\mathbf{T}^{(k)}$ and $\mathbf{M}^{(k)}$ are spherical tensor operators of rank $k$ in the electronic and nuclear space, respectively. The $k$ = 1 and $k$ = 2 terms represent the magnetic dipole and the electric quadrupole interactions. The higher-order hyperfine interactions, including the nuclear magnetic octupole interaction, are negligible~\cite{PhysRevA.77.012512,PhysRevA.87.012512}. The magnetic dipole and the electric quadrupole hyperfine interaction constants ($A$ and $B$) are defined as
\begin{equation}\label{eq:4}
A= \frac{ \mu_I \left \langle P J \left \| \mathbf{T}^{(1)} \right \| {PJ}\right \rangle}{I\sqrt{J(J+1)(2J+1)}}
\end{equation}
and
\begin{equation}\label{eq:5}
B=2Q\left \{\frac{J(2J-1)}{(J+1)(2J+3)} \right \}^{1/2}\left \langle PJ \left \| \mathbf{T}^{(2)} \right \| {PJ}\right \rangle.
\end{equation}
Here, $\mu_I$ and $Q$ are the magnetic dipole moment and the electric quadrupole moment of the nucleus. The electric tensor operators $\textbf{T}^{(1)}$ and $\textbf{T}^{(2)}$ are given by
\begin{equation}\label{eq:6}
\textbf{T}^{(1)} = \sum_{j}-i\alpha r_{j}^{-2}( \bm{\alpha}_j \cdot \mathbf{l}_j \mathbf{C}^{(1)}(j)),
\end{equation}
\begin{equation}\label{eq:7}
\textbf{T}^{(2)} = \sum_{j}-r_{j}^{-3}\mathbf{C}^{(2)}(j),
\end{equation}
where $i$ is the imaginary unit, $r_j$ is the radial coordinate of the $j$th electron, $\mathbf{l}$ is the orbital angular momentum operator, $\mathbf{C}^{(k)}$ is a spherical tensor of rank $k$, $\alpha$ is the fine structure constant and $\bm{\alpha}_j$ is the Dirac matrix.

According to Eq.~(\ref{eq:5}), the nuclear electric quadrupole moment (in mb) can be extracted by
\begin{equation}\label{eq:8}
Q = \frac{4.2559579}{\mathrm{EFG}} B,
\end{equation}
where the hyperfine interaction constant $B$ and the electric field gradient (EFG) which represents the electric charge distribution in the vicinity of the nucleus are expressed in MHz and a.u., respectively. In the present work, we will monitor the related $B/Q \propto$ EFG ratio value (in MHz/b) instead of the EFG.
\section{Calculations of $A$ and $B/Q$}
In this work, we adopted the active space approach to capture electron correlations~\cite{CPL48157,JCP892185}. According to the perturbation theory, the electron correlations can be divided into the first-order and the higher-order correlations~\cite{MCDHF1997,Li2012}. The first-order electron correlation effect, composed of valence-valence (VV), core-valence (CV) and core-core (CC) correlations, is captured by the CSFs generated through the single (S) and double (D) excitations from the occupied orbitals in the single reference configuration set. The higher-order electron correlations are usually considered in two ways~\cite{PhysRevA.96.012514}. One is to add CSFs generated by the triple (T) and quadruple (Q) excitations from the mono-reference configuration and the other is to include those produced through the SD excitations from the multi-reference (MR) configurations. By using the TQ-excitation method, the higher-order electron correlations are more difficult to deal with as the number of CSFs grows rapidly and easily goes beyond the current computer resources. Alternatively, the SD-excitation MR approach was adopted to estimate the higher-order electron correlations. In this method, the CSFs with large mixing coefficients in the first-order correlation were selected to form the MR configuration set. The CSFs generated by SD excitations from the MR configuration set can capture the main higher-order electron correlations. In this calculation, the outermost $5s$ and $5p$ electrons in the reference configuration were treated as the valence electrons and the others the core.
\subsection{Cd$^{+}$ ion}
For calculating the hyperfine interaction constants of the $5p~^2P_{1/2,3/2}$ states in Cd$^+$ ion, two different approaches were used. In the first approach, called optimal level 1 (OL1), the wave function for the $^2P_{3/2}$ state was optimized independently. In the second approach, called optimal level 2 (OL2), the wave function was generated with the EOL form. In the OL1 scheme, the energy of the $^2P_{3/2}$ state is optimized to minimize in variational calculations, but the OL2 scheme can expand the calculations to cover the $^2P_{1/2,3/2}$ states in the $5p$ configuration.

Our calculation was started in the Dirac-Hartree-Fock (DHF) approximation. The occupied orbitals in the reference configuration, or called spectroscopic orbitals, were optimized and kept frozen in the following computations. The VV and CV correlations between the electrons in the core and valence electron, labeled as CV model, were considered in the SCF procedure through the configuration space expanded by SD-excitation CSFs from the reference configuration $1s^22s^22p^63s^23p^63d^{10}4s^24p^64d^{10}5p$. The SD-excitation is restricted that at most one electron is excited from the core shells.
The occupied shells were successively opened for substitutions into the virtual set, starting with $4spd$, followed by $3spd$, and then by $2sp$ and $1s$. Meanwhile, the virtual orbital (VO) is added layer by layer, and only the added virtual orbitals were variable. In order to fully consider the VV and CV correlations, seven layers of VOs were augmented to make sure the convergence of parameters under investigation.
In Table~\ref{tab:1}, the progressively opened core shells and the expanded VOs are presented. The numbers of configuration state functions (NCSFs) for states in the $5p$ configuration generated in various computational models are also shown in this table. To raise computational efficiency, the CSFs which do not interact with the reference configurations were removed~\cite{Jonsson2007597,MCDHF1997}. The orbital set obtained in the CV model was used for the subsequent RCI calculation.
\begingroup
\squeezetable
\begin{table}
\caption{\label{tab:1}The numbers of configuration state functions (NCSF) in various computational models.}
\begin{ruledtabular}
\begin{tabular}{llcccc}
                                           &                               &         &          &   NCSF  &     \\ \hline
Reference configurations                   & VO                            & Model   &  $J$=1/2 &  $J$=3/2&   \\ \hline
\{$5p$\}                                   &                               &  DHF    &      1   &      1  &   \\
\{$4s^24p^64d^{10}5p$\}                    &\{5s, 6p, 5d, 4f\}             &         &    117   &    207  &   \\
\{$3s^2\cdots5p$\}                         &\{6s, 7p, 6d, 5f, 5g\}         &         &    877   &   1607  &  \\
\{$2s^2\cdots5p$\}                         &\{7s, 8p, 7d, 6f, 6g, 6h\}     &         &   2615   &   4849  &  \\
\{$1s^2\cdots5p$\}                         &\{8s, 9p, 8d, 7f, 7g, 6h\}     &         &   4816   &   8871  &  \\
\{$1s^2\cdots5p$\}                         &\{9s,10p, 9d, 8f, 8g, 6h\}     &         &   7428   &  13695  &  \\
\{$1s^2\cdots5p$\}                         &\{10s,11p,10d, 9f, 8g, 6h\}    &         &   9973   &  18305  &  \\
\{$1s^2\cdots5p$\}                         &\{11s,12p,11d, 9f, 8g, 6h\}    &  CV     &  11998   &  21853  &  \\
\{$4s^24p^64d^{10}5p$\}                    &\{11s,12p,11d, 9f, 8g, 6h\}    &  CC     &  67050   & 171725  & \\
\{$4s^24p^64d^{10}5p$\}                    &\{9s, 8p, 6d, 5f, 5g\}         &  Tri    &2442229   &4427350  &  \\
                                           &                               &         &  $J$=0   &   $J$=1 & $J$=2 \\
\{$5s5p$\}                                 &                               &  DHF    &      1   &      2  &      1     \\
\{$4s^24p^64d^{10}5s5p$\}                  &\{6s, 6p, 5d, 4f\}             &         &    246   &    871  &    787     \\
\{$3s^2\cdots5s5p$\}                       &\{7s, 7p, 6d, 5f, 5g\}         &         &   1793   &   7440  &   7229     \\
\{$2s^2\cdots5s5p$\}                       &\{8s, 8p, 7d, 6f, 6g, 6h\}     &         &   5285   &  23196  &  23024     \\
\{$1s^2\cdots5s5p$\}                       &\{9s, 9p, 8d, 7f, 7g, 6h\}     &         &   9677   &  43143  &  42946     \\
\{$1s^2\cdots5s5p$\}                       &\{10s,10p, 9d, 8f, 8g, 6h\}    &         &  14934   &  67678  &  67571     \\
\{$1s^2\cdots5s5p$\}                       &\{11s,11p,10d, 9f, 8g, 6h\}    &         &  20078   &  91403  &  90816     \\
\{$1s^2\cdots5s5p$\}                       &\{12s,12p,11d, 9f, 8g, 6h\}    &  CV     &  24178   & 109719  & 107940     \\
\{$4s^24p^64d^{10}5s5p$\}                  &\{12s,12p,11d, 9f, 8g, 6h\}    &  CC     &  58058   & 446978  & 433156     \\
+\{$4s^24p^64d^{8}5s5p5d^2$;               &\{8s, 7p, 6d, 4f\}             &  MR     &2021707   &6405941  &9432299     \\
$4s^24p^64d^{10}5p6s$;                     &                               &         &          &         &            \\
$4s^24p^64d^{10}5p5d$;                     &                               &         &          &         &            \\
$4s^24p^64d^{10}5s6d$\}                    &                               &         &          &         &              \\
\end{tabular}
\end{ruledtabular}
\end{table}
\endgroup

The CC electron correlation related to the $n = 4$ shell was estimated with the CSFs generated by SD excitations from the $n = 4$ core shell to the full layers of the virtual orbitals. This computational model is marked as CC in Table~\ref{tab:1}. As analyzed in Ref.~\cite{Porsev2006,0953-4075-41-11-115002,PhysRevA.79.052502}, the three-particle effects are often very important for the hyperfine interactions. Therefore, the contributions of the higher-order electron correlation effects were considered by triple excitations from the $n =4, 5$ core shell in this work. The CSF space is further expanded by triple excitations from the single reference configuration to the first three layers of the virtual orbitals, which is labeled as `Tri'. In practice, the GRASP2018 package~\cite{Fischer2019} was employed to perform calculations.

\begin{figure}
 \includegraphics[scale=0.9]{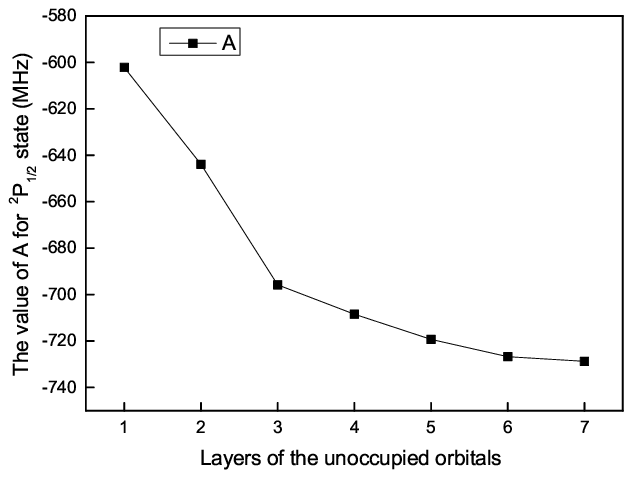}
 \includegraphics[scale=0.9]{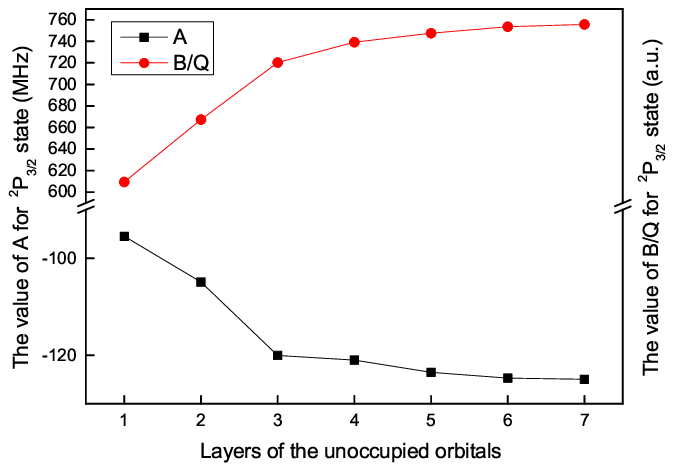}
 \caption{\label{fig:1}The values of $A$ constants and $B/Q$ for $^2P_{1/2}$ and $^2P_{3/2}$ states as functions of the virtual layers in the CV model (OL2 approach).}
\end{figure}

In Table~\ref{tab:2}, the values of $A$ and $B/Q$ for $^2P_{1/2}$ and $^2P_{3/2}$ states in $^{109}$Cd$^+$ ion in various computational models are shown. Previous experimental measurement of the hyperfine interaction constant $A$ and theoretical calculation of the $B/Q$ have also been displayed. From the table, we can see that the calculated results from the OL1 and OL2 approaches are almost the same. It also can be seen that the VV and CV correlations are the dominant correlation effects for the hyperfine parameters. In Fig.~\ref{fig:1}, the values of $A$ constants and $B/Q$ for the $^2P_{1/2}$ and $^2P_{3/2}$ states in the OL2 approach as functions of the virtual layers are presented. From the figures, it can be seen that the VV and CV correlation effects on the hyperfine interaction constants are converged when seven virtual layers were added. The CC correlation in the $n = 4$ core shell is also important. It contributes about 24\% for the hyperfine constants $A$ of the $^2P_{1/2}$ state and 27\% for the $^2P_{3/2}$ state. For $B/Q$, its contribution is about 26\% for the $^2P_{3/2}$ state.

As is well known that three-particle effects are important for the hyperfine structure~\cite{PhysRevA.97.062505,PhysRevA.53.4021}. To consider the higher-order electron correlations, we adopted the triple substitutions from the $n = 4, 5$ shells. It can be observed that the contribution of the triple substitutions on the hyperfine interaction constants are about 6\%, opposite to the contribution of the CC correlation. The calculated $A$ value for the $^2P_{3/2}$ state obtained in the `Tri' model is closer to the experimental measurement. The four-particle effects on the values of $A$ and $B/Q$ were estimated to be about 0.2\% and 0.1\%, which determined that the higher-order electron correlations considered by the triple substitutions are sufficient. Therefore, the CC and higher-order electron correlations cannot be neglected for accurately calculating the hyperfine parameters in Cd$^+$. In addition, it was found that the Breit interaction and QED corrections have negligible effects on the hyperfine structures.

Our calculated value of $A$ for the $^2P_{3/2}$ state agrees to about the 6\% level the experimental measurement~\cite{Yordanov2013}. This rough agreement determined that our computational models are reasonable. The error bar of the magnetic dipole hyperfine interaction constant $A$ can be used as a source for estimating the uncertainty of the calculated $B/Q$, since they both have similar $r^{-3}$ dependence on the radial part of the electronic wave function~\cite{PhysRevA.71.012502, PhysRevLett.87.133003}. Additionally, our calculated value of $B/Q$ for the $^2P_{3/2}$ state lies in the range of the result with error bar given by Yordanov~\textit{et al.}~\cite{Yordanov2013} using the Dirac-Hartree-Fock calculation.
\begin{table}
\caption{\label{tab:2}$A$ (MHz), $B/Q$ (MHz/b) and $Q$ (b) values for the $^2P_{1/2}$ and $^2P_{3/2}$ states in $^{109}$Cd$^+$ ($I^{\pi}$ = 5/2$^+$ and $\mu$ = $-$0.8278461(15)$\mu_N$~\cite{Froemmgen2015}) calculated with OL2 and OL1 (in parentheses) approaches in various computational models. The $Q$ values are extracted from the relation $Q = B_{expt}/(B/Q)$, where the experimental value is $B_{expt}(^2P_{3/2})$ = 403(1)~\cite{Yordanov2013}.}
\begin{ruledtabular}
\begin{tabular}{ccccc}
                                      &       $^2P_{1/2}$     &     \multicolumn{3}{c}{$^2P_{3/2}$}                   \\  \cline{3-5}
Models                                &        $A$ (MHz)      &    $A$ (MHz)       &  $B/Q$ (MHz/b)  &  $Q$ (b)        \\\hline
DHF                                   &        $-$506.23      &  $-$78.63($-$78.54)    & 454.85(454.33)  & 0.886(0.887)    \\
CV                                    &        $-$728.72      &  $-$124.96($-$124.58)  & 755.72(754.57)  & 0.533(0.534)    \\
CC                                    &        $-$586.82      &  $-$98.08($-$97.71)   & 599.81(598.40)  & 0.672(0.673)    \\
Tri                                   &        $-$613.76      &  $-$104.52($-$104.28)  & 640.94(640.76)  & 0.629(0.629)    \\
Others                                &                       &  $-$111.4(2)\footnotemark[1]& 666(27)\footnotemark[2]  &                 \\
\end{tabular}
\footnotetext[1]{Measured by Yordanov~\textit{et al.}~\cite{Yordanov2013}}
\footnotetext[2]{Calculated by the Dirac-Hartree-Fock calculation~\cite{Yordanov2013}}
\end{ruledtabular}
\end{table}
\subsection{Cd atom}
Similar to the calculations in Cd$^+$, two different approaches were also used for the hyperfine structure calculations for the $5s5p~^3P_{1,2}$ states in neutral Cd atom. In the first approach (OL1), the wave functions for the $^3P_1$ and $^3P_2$ states were optimized independently. In the second approach (OL4), the wave functions were optimized for all four states of the $5s5p$ configuration ($5s5p~^3P_0,^3P_1,^3P_2,^1P_1$) in the EOL approach, with standard weights.

After the calculation in the DHF approximation, the VV and CV correlations were considered in the SCF procedure through the configuration space expanded by restricted SD-excitation CSFs from the $5s5p$ configuration. The restricted double excitations were described as the same as the calculations in Cd$^+$. And the occupied shells in the reference configuration were opened in succession while the virtual orbital is added layer by layer. The reference configuration(s), the VOs and the NCSFs for states in the $5s5p$ configuration generated in different computational models are also presented in Table~\ref{tab:1}.

Furthermore, the CC correlation in the $n = 4$ core shell was included as well in the RCI calculation. In this calculation, the configuration space was expanded with SD excitations from the occupied $n$ = 4 core shell to the full layers of virtual orbitals. The MR-SD model was applied to estimate the higher-order electron correlation effects among the $n$ = 4, 5 shells. As mentioned earlier, the MR configuration set was formed by selecting the dominant CSFs in the CC model, i.e., those CSFs with mixing coefficients $c_j$ larger than 0.04. In this work, the MR-configuration is composed of \{$4s^24p^64d^85s5p5d^2$; $4s^24p^64d^{10}5p6s$; $4s^24p^64d^{10}5p5d$; $4s^24p^64d^{10}5p6d$\}. The SD excitations were allowed from the MR configurations to three layers of virtual orbitals. This model is marked as MR.

\begin{figure}
 \includegraphics[scale=0.9]{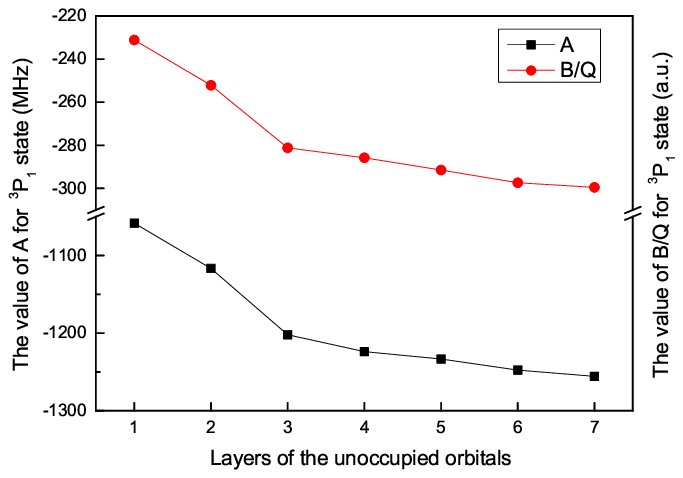}
 \includegraphics[scale=0.9]{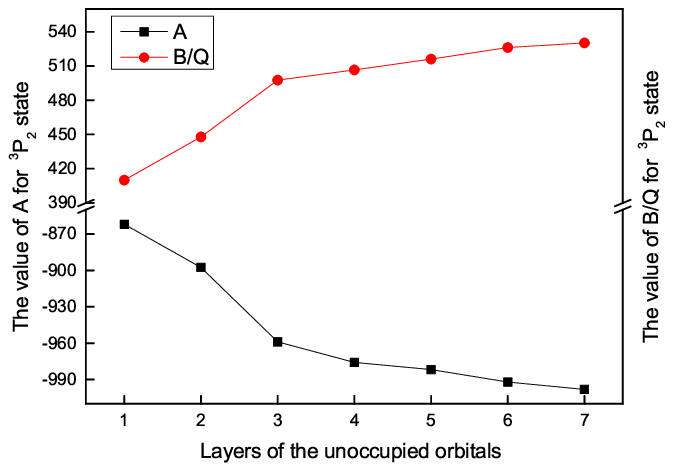}
 \caption{\label{fig:2}The values of $A$ constants and $B/Q$s for $^3P_1$ and $^3P_2$ states as functions of the virtual layers in the CV model (OL4 approach).}
\end{figure}

The values of $A$ and $B/Q$ for the $5s5p~^3P_{1,2}$ states in $^{109}$Cd atom in various computational models are shown in Table~\ref{tab:3}. The previous experimental measurement of the hyperfine interaction constant $A$ and theoretical calculation of the $B/Q$ have been displayed as well. From the table, we can see that there is a difference between the calculated results from the two approaches (OL1 and OL4). For constant $A$, the difference is less than 1\%, but about 6\% for $B/Q$. The VV and CV correlations make the main contribution to the hyperfine interaction constants of both states. In Fig.~\ref{fig:2}, we present the values of $A$ constants and $B/Q$s for the $^3P_1$ and $^3P_2$ states in the OL4 approach as functions of the virtual layers. After optimizing seven layers of correlation orbitals, the VV and CV correlation effects on the hyperfine interaction constants were converged. The CC correlation in the $n = 4$ core shell changes the $A$ constants by a factor of 19\% for the $^3P_1$ and 14\% for the $^3P_2$ states, and 37\% for the $B/Q$ values of the two states. We can also observe that the higher-order electron correlations, estimated in the MR-SD model, make an opposite contribution to the CC correlation effects, which was also observed in Ref.~\cite{PhysRevA.100.012504} for Sr I, Ref.~\cite{PhysRevA.79.052502} for Au I and Ref.~\cite{PhysRevA.96.012514} for the Al$^+$ ion. Therefore, the CC correlation and the higher-order electron correlations cannot be neglected for estimating the uncertainty. To the best of our knowledge, there is no experimental value for the $^1P_1$ state, so we exclude this state as a reference to extract the $Q$ value.

Fr$\rm{\ddot{o}}$mmgen~\textit{et al.}~\cite{Froemmgen2015} also calculated the values of $A$ and $B/Q$ for the $^3P_2$ state by using the MCDHF method. By following the calculation model presented in Ref.~\cite{Froemmgen2015}, our calculated value of $A$ constant obtained in the CV model agrees with theirs, but the $B/Q$ value is about 9\% larger. The reason for this difference is unclear at present. Our values of $A$ for $^3P_1$ and $^3P_2$ states in the final model are in close agreement with the experimental measurements. This determined that our computational model is reasonable.
\begin{table}
\caption{\label{tab:3}$A$ (MHz), $B/Q$ (MHz/b) and $Q$ (b) values for the $5s5p~^3P_{1,2}$ states in $^{109}$Cd ($I^{\pi}$ = 5/2$^+$ and $\mu$ = $-$0.8278461(15)$\mu_N$~\cite{Froemmgen2015}) calculated with OL4 and OL1 (in parentheses) approaches in various computational models. The $Q$ values are extracted from the relation $Q = B_{expt}/(B/Q)$, where the experimental values are $B_{expt}(^3P_1)$ = $-$165.143(5)~\cite{Phys.Rev.132.1186} and $B_{expt}(^3P_2)$ = 296(2)~\cite{Froemmgen2015}.}
\begin{ruledtabular}
\footnotesize
\begin{tabular}{ccccccc}
Models       &       \multicolumn{3}{c}{$^3P_1$}                            & \multicolumn{3}{c}{$^3P_2$} \\ \cline{2-4} \cline{5-7}
             &  $A$ (MHz)            & $B/Q$ (MHz/b)        & $Q$ (b)       & $A$ (MHz)            & $B/Q$ (MHz/b)  & $Q$ (b)     \\\hline
DHF          & $-$884.04($-$882.04)  & $-$156.64($-$191.53) & 1.054(0.862)  & $-$761.48($-$735.71) & 272.28(334.86) & 1.087(0.884) \\
CV           & $-$1255.68($-$1258.46)& $-$299.52($-$306.47) & 0.551(0.539)  & $-$998.23($-$998.56) & 530.31(545.83) & 0.558(0.542) \\
CC           & $-$1057.17($-$1056.36)& $-$218.90($-$251.13) & 0.754(0.658)  & $-$876.28($-$860.49) & 385.86(446.97) & 0.767(0.662) \\
MR           & $-$1118.03($-$1120.47)& $-$254.77($-$266.15) & 0.648(0.620)  & $-$910.06($-$904.97) & 452.06(474.03) & 0.655(0.624) \\
Others       & $-$1148.784(7)\footnotemark[1] &       &  &  $-$917.5(1)\footnotemark[2]     & 500(30)\footnotemark[3] &           \\
\end{tabular}
\footnotetext[1]{Measured by Thaddeus and Mcdermott~\cite{Phys.Rev.132.1186}}
\footnotetext[2]{Measured by Fr$\rm{\ddot{o}}$mmgen~\textit{et al.}~\cite{Froemmgen2015}}
\footnotetext[3]{Calculated by the MCDHF method~\cite{Froemmgen2015}}
\end{ruledtabular}
\end{table}
\section{The nuclear electric quadrupole moment of $^{109}$C\MakeLowercase{d}}
Combining the calculated $B/Q$ of the $^2P_{3/2}$ state in Cd$^+$ ion and the $^3P_1$, $^3P_2$ states in Cd atom with the measured hyperfine interaction constants $B$($^2P_{3/2}$) = $-$403(1)~MHz~\cite{Yordanov2013}, $B$($^3P_1$) = $-$165.143(5)~MHz~\cite{Phys.Rev.132.1186} and $B$($^3P_2$) = 296(2)~MHz~\cite{Froemmgen2015}, six values of the nuclear electric quadrupole moment $Q$($^{109}$Cd) were obtained. The estimated results are also presented in Tables~\ref{tab:2} and~\ref{tab:3}.

The scattering of these six values gives us an information about the dependence of the calculated values of $B/Q$ on the choice of our computational methods and provides a basis for estimating the error bar of the nuclear quadrupole moment $Q$($^{109}$Cd). The estimated values of $Q$ are distributed around their median value ($Q$ = 0.629~b), with a rather flat distribution. For an estimate of the accuracy, we obtained a value of $Q$ = 0.634~b on average of the six evaluations with one standard deviation of 14~mb. The difference between the median and average results is negligible compare to the error bar.

As mentioned above, the difference between the calculated and experimental values of the magnetic dipole hyperfine constant $\left| A_{Cal.} - A_{Exp.} \right|$ may be used as another tool to estimate the error bar for the determination of $Q$~\cite{PhysRevA.79.052502,PhysRevA.97.062505,PhysRevA.100.012504,Schiffmann2020}. Considering the uncertainties of $A$ for the $^2P_{3/2}$, and $^3P_{1,2}$ states, an average error bar of 22~mb is obtained. We chose the larger of the above two error bars and finally proposed that $Q$($^{109}$Cd) = 0.634(22)~b.

Table~\ref{tab:4} shows the available nuclear electric quadrupole moments of the $^{109}$Cd nucleus for comparison. To our knowledge, Laulainen and McDermott~\cite{Laulainen1969} first reported a nuclear electric quadrupole moment of $^{109}$Cd, based on their measurement on the hyperfine-structure splitting of the $5s5p~^3P_1$ state by the optical double-resonance technique and a simple calculation on the EFG in the semi-empirical approximation. This value has been used as a reference for almost 50 years~\cite{Stone2005}. Yordanov~\textit{et al.}~\cite{Yordanov2013} calculated the value of $B/Q$ for the $5p~^2P_{3/2}$ state in $^{109}$Cd$^{+}$ ion by using the Dirac-Hartree-Fock calculation. The $Q$ value was extracted from the combination of the measurement of $B$ using the collinear laser spectroscopy and the calculation of $B/Q$. This value was mentioned in the review from Pyykk\"{o}~\cite{Pyykk2017} and Stone~\cite{Stone2016}. Most recently, Fr$\rm{\ddot{o}}$mmgen~\textit{et al.}~\cite{Froemmgen2015} also reported a value $Q$ = 0.592(36) based on the $B/Q$ value of the $5s5p~^3P_2$ state in the $^{109}$Cd atom given by the MCDHF theory. This value matches the quoted value of $Q$ = 0.604(25)~b. Only the main first-order electron correlations were included in their calculations, and the corrections resulting from unrestricted double and triple substitutions were assessed to be smaller than the uncertainty of $B/Q$. Compared to the two previous values, our value is about 5\% larger and the error bar is relatively smaller. It should be stressed that the nuclear quadrupole moment of $^{109}$Cd nucleus was extracted from the hyperfine structures of atomic and ionic cadmium. The values of $B/Q$ were calculated in different theoretical frameworks with detailed consideration of the electron correlation effects, especially the CC and the higher-order electron correlations.
\begin{table}
\caption{\label{tab:4}$Q$ value from various works in barn (b).}
\begin{ruledtabular}
\begin{tabular}{ccc}
                                                            &  $Q$(b)     &  Atomic systems \\ \hline
Laulainen and McDermott~\cite{Laulainen1969}                & 0.69(7)     &   $5s5p~^3P_1$~$^{109}$Cd atom  \\
Yordanov~\textit{et al.}~\cite{Yordanov2013}                & 0.604(25)   &   $5p~^2P_{3/2}$~$^{109}$Cd$^+$ ion  \\
Fr$\rm{\ddot{o}}$mmgen~\textit{et al.}~\cite{Froemmgen2015} & 0.592(36)   &   $5s5p~^3P_2$~$^{109}$Cd atom \\
This work                                                   & 0.634(22)   &   $5p~^2P_{3/2}$~$^{109}$Cd$^+$ ion \\
                                                            &             &   $5s5p~^3P_{1,2}$~$^{109}$Cd atom \\
\end{tabular}
\end{ruledtabular}
\end{table}
\section{Conclusion}
By using the MCDHF theory, six different computational approaches were employed to determine the expectation values of the $B/Q$s for the $5p~^2P_{3/2}$ state in $^{109}$Cd$^+$ ion and the $5s5p~^3P_{1,2}$ states in neutral $^{109}$Cd atom. Different electron correlation effects on the hyperfine interaction constants $A$ and the $B/Q$s, especially for the correlations related to the core shells and higher-order electron correlations, were considered systematically. It was found that the contributions of the CC and the higher-order electron correlations on the $B/Q$s are remarkable and offset each other partly. Therefore, it is essential to take into account both of them in order to achieve high accuracy of hyperfine parameters. Combining the calculated $B/Q$s with the precise experimental values of the electric quadrupole hyperfine interaction constant $B$, we determined the nuclear quadrupole moment of $^{109}$Cd, $Q$ = 0.634(22)~b. Our result was obtained based on a combination of neutral atomic and single ionic systems. We obtained a reliable uncertainty for the $Q$ value which was controlled at 3\% level by accounting for statistical principles and the correlation with the magnetic dipole hyperfine interaction constant $A$. The nuclear electric quadrupole moment of $^{109}$Cd isotope can be used as a reference for determining the quadrupole moment of the other isotopes of Cd, especially the Cd isomers.

\begin{acknowledgments}
This work is supported by the National Natural Science Foundation of China under Grant No. 61775220, the Strategic Priority Research Program of the Chinese Academy of Sciences under Grant No. XDB21030100, the Key Research Project of Frontier Science of the Chinese Academy of Sciences under Grant No. QYZDB-SSW-JSC004, and the West Light Foundation of the Chinese Academy of Sciences under Grant No. XAB2018B17.
\end{acknowledgments}

\bibliography{mylib}

\end{document}